\documentclass[12pt,preprint]{aastex}
\newcommand  \ergs     {\ifmmode {\rm ergs\,s}^{-1} \else ergs s$^{-1}$\fi}
\newcommand  \ergcms   {\ifmmode {\rm ergs\,cm}^{-2}\,{\rm s}^{-1}
                        \else ergs\,cm$^{-2}$\,s$^{-1}$\fi}
\def\Msun{\ifmmode M_{\odot} \else $M_{\odot}$\fi}
\def\Lsun{\ifmmode L_{\odot} \else $L_{\odot}$\fi}

\slugcomment{Accepted for publication in ApJ main journal}

\shorttitle{Star formation in the hosts of high-z QSOs}
\shortauthors{Lutz et al.}

\begin{document}

\title{Star formation in the hosts of high-z QSOs: Evidence from Spitzer 
PAH detections}

\author{D. Lutz\altaffilmark{1}, 
E. Sturm\altaffilmark{1}, L.J. Tacconi\altaffilmark{1}, E. Valiante\altaffilmark
{1}, M. Schweitzer\altaffilmark{1}
H. Netzer\altaffilmark{2},
R. Maiolino\altaffilmark{3},
P. Andreani\altaffilmark{4,5},
O. Shemmer\altaffilmark{6},
S. Veilleux\altaffilmark{7}}
\altaffiltext{1}{Max-Planck-Institut f\"ur extraterrestrische Physik,
Postfach 1312, 85741 Garching, Germany \email{lutz@mpe.mpg.de,
sturm@mpe.mpg.de, linda@mpe.mpg.de,
valiante@mpe.mpg.de, schweitzer@mpe.mpg.de}}
\altaffiltext{2}{School of Physics and Astronomy and the Wise Observatory,
      The Raymond and Beverly Sackler Faculty of Exact Sciences,
     Tel-Aviv University, Tel-Aviv 69978, Israel\email{netzer@wise1.tau.ac.il}}
\altaffiltext{3}{INAF, Osservatorio Astronomico di Roma, via di Frascati 13,
   00040 Monte Porzio Catone, Italy\email{maiolino@ao-roma.inaf.it}}
\altaffiltext{4}{ESO, Karl-Schwarzschildstra\ss{}e 2, 85748 Garching, Germany
  \email{pandrean@eso.org}}
\altaffiltext{5}{INAF, Osservatorio Astronomico di Trieste, via Tiepolo 11,
  34143 Trieste, Italy}
\altaffiltext{6}{Department of Astronomy and Astrophysics, 525 Davey
Laboratory, Pennsylvania State University,
   University Park, PA 16802, USA\email{ohad@astro.psu.edu}}
\altaffiltext{7}{Department of Astronomy, University of Maryland,
      College Park, MD 20742-2421, USA\email{veilleux@astro.umd.edu}}

\begin{abstract}
We present Spitzer rest-frame mid-infrared spectroscopy of twelve z$\sim$2 
mm-bright type 1 QSOs, selected from unlensed and lensed QSO samples and 
covering a range of AGN optical luminosities L$_{5100}=10^{45}$ to 
10$^{47}$ erg s$^{-1}$.
On top of the AGN continuum, we detect PAH emission from luminous star 
formation in nine objects individually as well as in the composite spectrum 
for the full sample. PAH luminosity and rest frame far-infrared luminosity 
correlate and
extend the similar correlation for lower luminosity local QSOs. This provides 
strong evidence for intense star formation in the hosts of these mm-bright
QSOs, sometimes exceeding 1000\Msun yr$^{-1}$ and dominating their rest frame 
far-infrared emission. The PAH-based limit on star formation rates is lower 
for luminous z$\sim$2 QSOs that are not preselected for their mm emission. 
Partly dependent on systematic changes 
of the AGN dust covering factor and the dust spectral energy distribution of
the AGN proper, the spectral energy distributions of mm-faint high-z QSOs 
may be AGN 
dominated out to rest frame far-infrared wavelengths. Towards the most 
luminous high-z QSOs, there is a flattening of the 
relation between star formation and AGN luminosity that is observed for 
lower redshift QSOs. No QSO in our sample has a PAH-measured
star formation rate in excess of 3000 \Msun yr$^{-1}$.
 
\end{abstract}


\section{Introduction}
Studies of the evolution of galaxies and active galactic nuclei (AGN) are 
drawing an 
increasingly detailed picture of their global evolution from redshifts 
above 6 to the present. The cosmic star forming density as well as the number
density of AGN rise and fall over cosmic time, with a maximum in the `quasar 
epoch' at redshifts around 2. Less is known about the connection of star
formation and nuclear activity in individual high redshift objects, and
the controlling physical processes. A related basic observational goal is 
to quantify both AGN luminosity and star formation in an as wide as possible
set of galaxy populations, a task that is posing considerable technical 
difficulties for some types of objects. 

A particularly interesting population are the luminous AGN that are 
traced by optically selected QSOs.
Evolutionary connections between intense starburst events and QSOs have 
been proposed
early on \citep[e.g.][]{sanders88,norman88} and naturally suggest that the
luminous AGN phase coincides with or follows after a period of 
intense star formation. While still limited in spatial resolution and requiring
simplifying assumptions, increasingly advanced models of these processes 
now become available. They combine galaxy merging, gas transport, star 
formation, accretion and feedback 
processes \citep[e.g.][]{springel05}. In turn, such scenarios are part 
of recent global modes of galaxy and merger evolution 
\citep[e.g.][]{granato04,hopkins06}.

Indications have been obtained for intense star formation events in the host
of high redshift QSOs. Perhaps the most important is the detection of
submm or mm continuum (rest frame far-infrared) from dust in part 
of the high-z QSO population. If heated by star formation, this 
emission indicates
high star formation rates similar to those in submillimeter galaxies
\citep[e.g.][]{omont03,priddey03,barvainis02}. Large molecular gas reservoirs
sufficient to sustain intense star formation have been detected
in CO \citep[see summaries in][]{solomon05,greve05},
and in some of the brightest systems also in tracers of dense molecular gas 
\citep{barvainis97,solomon03,carilli05,riechers06,gracia06,guelin07}.
Finally, the [CII] 157$\mu$m fine structure line
was detected in a few high-z quasars \citep{maiolino05,iono06} at a ratio 
to the rest frame far-infrared emission similar to the ratio in local
ULIRGs, consistent with massive star formation. While these observations
indicate a plausible picture of massive star formation in some QSO hosts, they
need independent support, in particular to better discriminate between star 
formation and AGN heating of the observed rest frame far-infrared continuum. 
We have hence initiated a program using the mid-infrared `PAH' emission 
features as tracers of star formation in the hosts of high-z QSOs.
  
Because of their presence in star forming environments with a wide range of
physical conditions but absence close to powerful AGN, the main mid-infrared
PAH emission features at 6.2, 7.7, 8.6, and 11.3$\mu$m have been used 
extensively as extragalactic star formation indicators. The calibration 
factor between PAH luminosity and star formation rate
depends on average physical conditions in the interstellar medium of the
galaxy probed. This is indicated both
in the relatively weaker PAH in galaxies with intense interstellar 
radiation fields \citep[as measured by the large grain dust temperature from 
their SEDs,][]{dale01}, and in the sublinear increase of PAH emission 
with P$\alpha$
hydrogen recombination line emission, that is observed in spatially resolved 
studies of nearby 
galaxies \citep{calzetti07}. For the specific case of star formation 
indicators for QSO hosts, the associated uncertainty is outweighed by several 
advantages. PAH emission from starbursts is very luminous and thus 
less easily outshone by the powerful AGN. Unlike other popular star 
formation indicators like H$\alpha$ or mid-infrared 24$\mu$m 
continuum, it will not be quickly overwhelmed by strong AGN emission in the 
very same tracers. It is also less sensitive than e.g. the optical 
[OII] 3727\AA\ line
to the  significant internal obscuration that is typical for starburst 
regions. 

In \citet{schweitzer06} and \citet{netzer07} we have applied these techniques
to local z$\leq$0.3 Palomar-Green (PG) QSOs, detecting PAH emission in many
objects and concluding that likely most of their 
$\sim 10^{10}$--10$^{12}$\Lsun\ far-infrared luminosity is due to star 
formation. Comparing average spectra and SEDs of groups of QSOs that are 
differing in their level of far-infrared emission, we were able to 
derive an infrared SED for the
pure AGN, after subtraction of host star formation.

Several spectroscopic studies have shown that Spitzer-IRS spectroscopy with
modest integration times of at most a few hours per source can successfully
detect the PAH emission of the most luminous z$\sim$2.5 star forming events, 
which approach infrared luminosities of 10$^{13}$\Lsun\ and star 
formation rates around 1000\Msun /yr. Such detections have been obtained for
 submillimeter 
galaxies (SMGs) \citep{lutz05,valiante07,menendez07,pope08} as well as for 
a subset of bright Spitzer 24$\mu$m selected galaxies 
\citep[e.g.][]{weedman06,yan07}. Substantially 
longer integrations exceeding 10 hours in an IRS spectral order have been 
demonstrated to be able to detect yet fainter PAH emission at similar 
redshifts with a peak of the rest frame 7.7$\mu$m feature of 0.4mJy or less
\citep{teplitz07}. Probing PAH emission to these levels would be highly 
interesting for 
a complete sample of luminous z$\sim$2.5 QSOs but expensive in observing
time. Systematic difficulties arise in detecting PAH features on a strong AGN 
continuum that 
additionally has structure due to silicate emission. The silicate emission
is strong in regions of the spectrum
where the main PAH complexes are weak, and thus tends to wash out the 
PAH-induced structure, as demonstrated for some local PG QSO spectra in 
\citet{schweitzer06}.

In this paper, we hence focus as an important first step on 
measuring PAH emission and inferred star 
formation in a modest size sample of twelve z=1.8 to z=2.8  QSOs that all 
have detected 
(sub)mm emission. If this mm emission is due to host star formation,
its associated PAH emission will be detectable with IRS in modest integration
times by analogy to the SMGs at similar redshift. In \citet{lutz07} we have 
presented first results for the brightest object in our sample, the Cloverleaf
(H1413+117). Here we present the full mm-bright sample and discuss in 
section 4.2 implications for the 
high redshift QSO 1 population as a whole.

Throughout the paper we adopt an $\Omega_m =0.3$, $\Omega_\Lambda =0.7$ 
and $H_0=70$ km\,s$^{-1}$\,Mpc$^{-1}$ cosmology. When converting IR 
luminosities to star formation rates we adopt for consistency
with much of the previous work the \citet{kennicutt98} value
which is based on a 0.1-100\Msun\ Salpeter IMF. For a more realistic Chabrier
or Kroupa IMF star formation rates would be a factor $\sim$1.6 smaller. Our 
sample includes lensed sources. For these,
we present fluxes or flux densities as observed, but luminosities, star 
formation rates and 
masses corrected for the adopted lensing magnification.

\section{Sample and Observations}

We have selected from the literature z$\sim$2 Type 1 radio quiet QSOs with 
robust ($>5\sigma$) and consequently bright submm or mm dust continuum 
detections. The 
sample includes lensed as well as unlensed QSOs and is summarized in 
Table~\ref{tab:sample}. Sources have been chosen from the surveys of (sub)mm
emission in z$\sim$2 QSOs by \citet{omont03} and \citet{priddey03}, the lensed
QSO submm survey of \citet{barvainis02}, and one target each from 
\citet{guilloteau99} and \citet{knudsen03}. One of the targets 
(SMMJ04315+10277, \citet{knudsen03}) was first reported in a submm survey 
and may be
intermediate in properties between optically selected QSOs and typical submm 
selected
galaxies. The other targets are selected from the mentioned studies of 
optically selected QSOs solely
on the basis of a bright and robust (sub)mm detection and will provide a view
of the rest frame far-infrared bright part of the general z$\sim$2 unobscured 
QSO population.
Our unlensed targets have AGN rest frame optical luminosities 
$L_{5100}= \nu L_\nu(5100\AA)$ in the range 10$^{46}$ to 10$^{47}$\ergs, some
of the lensed QSOs are intrinsically an order of magnitude less luminous
and overlap with the highest luminosity local sources studied with
similar methods by \citet{schweitzer06} and \citet{netzer07}.

For targets in \citet{omont03} the optical luminosities were adopted from 
the rest frame blue absolute magnitudes in this reference, converting 
to our adopted cosmology. In the other cases we used published near-infrared 
photometry \citep[][and 2MASS point source catalog]{priddey03}, and the 
assumption of a $f_\nu\propto\nu^{-0.44}$ rest frame optical/UV continuum
\citep{vandenberk01}.

We obtained low resolution (R$\sim 60-120$) mid-infrared spectra of the QSOs 
using the Spitzer infrared spectrograph IRS \citep{houck04} during the period
July 2006 - October 2007. The IRS LL slit is wide enough to include
all images of the lensed sources. For each target, 30 cycles of
120 sec integration time per nod position were taken in the
LL1 (19.5 to 38.0 $\mu$m) module and 15
cycles in the LL2 (14.0 to 21.3 $\mu$m) module, leading to effective on-source
integration times of 2 and 1 hours, respectively. We start from the pipeline
processed basic calibrated data (version 14.4 to 16.1 for different sources),
subtracting individually data for the same data collection 
event within the data collection sequences of the two nod positions. 
We use own deglitching routines identifying outliers from the default bad 
pixel masks, by identifying points discrepant by more than a multiple of 
the local noise from their spectral neighborhood, and by visual inspection 
of the 2D frames, finally replacing them with values representative of 
their spectral neighborhood. We coadded the images for each cycle,
checking pixel per pixel for outlying values in the stack, and subtracted 
residual background derived from a first order spatial fit to source-free 
regions. SMART \citep{higdon04} and SPICE (optimum extraction,
for the weak continuum source SMMJ04315 only) were used for 
extraction. We compared the independent spectra from the two nod positions 
before averaging, finding no mismatches that would be indicating 
mispointings or other serious problems with the data.  Before combining 
the segments of the spectrum observed in the 
different modules into the final spectrum, the segment observed in the LL2 
module was multiplied with an object-dependent scaling factor (within a few \% 
of 1) in order to provide best match in the overlapping wavelength range. 

\section{Results}
\begin{figure}
\epsscale{1.}
\plotone{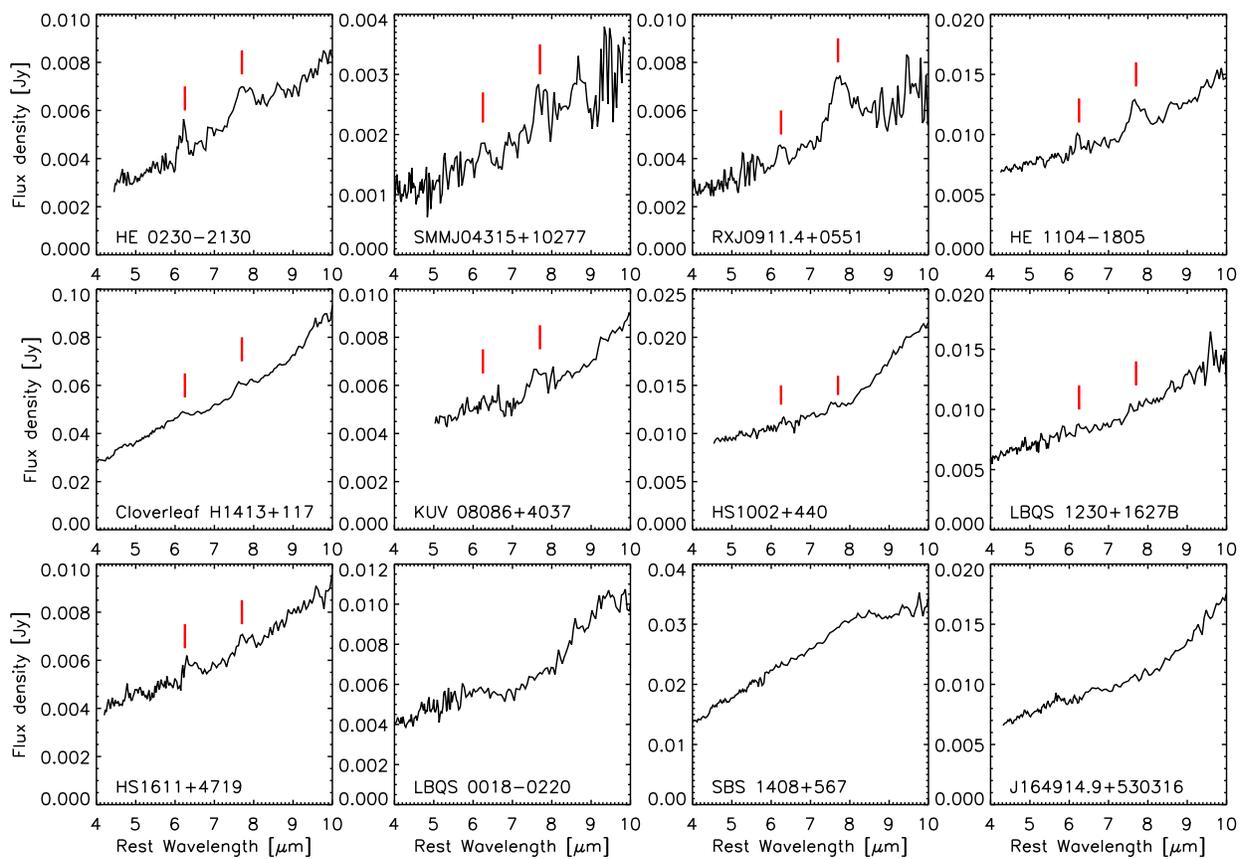}
\caption{Spitzer-IRS spectra of the sample of (sub)mm-bright high-z QSO1s. 
Vertical lines indicate the location of the 6.2 and 7.7$\mu$m PAH features at
the redshift of the sources with PAH detections.}
\label{fig:allhighz}
\end{figure}

\begin{figure}
\epsscale{.70}
\plotone{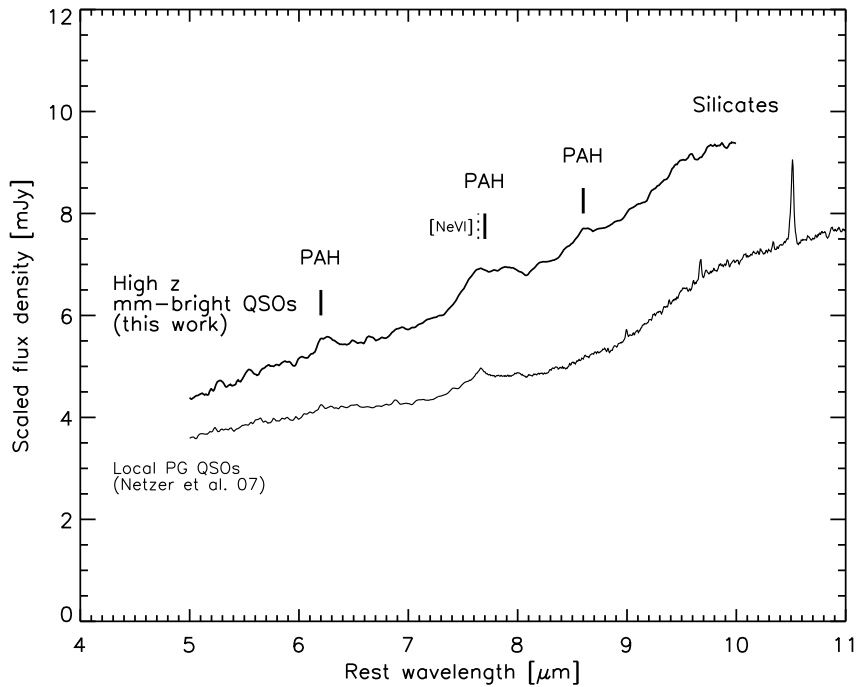}
\caption{Average rest frame mid-IR spectrum of the sample high-z QSO1s. The 
individual spectra have been scaled before averaging to the same flux of 
15mJy at rest frame wavelength 222$\mu$m. This allows a direct comparison
of spectrum and the relation of PAH and far-infrared emission to the stacked 
submm galaxy spectrum 
presented by \citet{valiante07}. A direct average of the QSO spectra has
a very similar shape. A
scaled average spectrum of the local PG QSOs studied by \citet{netzer07} is 
shown for comparison.}
\label{fig:averagehighz}
\end{figure}

Figure~\ref{fig:allhighz} shows the spectra of all our targets, and 
Fig.~\ref{fig:averagehighz} a composite derived from all sample objects.
The main spectral features are the ones also seen in the spectra of local PG
QSOs \citep[e.g.][]{siebenmorgen05,hao05,schweitzer06,hao07,netzer07}: A 
strong AGN continuum
with superposed silicate emission at $\lambda\gtrsim 9\mu$m, and aromatic 
`PAH' emission features at 6.2, 7.7, 8.6$\mu$m at varying strength 
relative to the local continuum. An unusual individual spectrum is the one
of SBS 1408+567 with convex continuum, a smooth broad emission feature
around 8$\mu$m, but no unambiguous identification of PAH emission. This broad
feature is weaker relative to the continuum but has similarity with the
8$\mu$m feature in the mid-infrared spectrum of the well known Type 1 AGN
and ULIRG 
Mrk 231 \citep{rigopoulou99,spoon02,weedman05,armus07} which may originate
from a combination of silicate emission and absorption. PAH 7.7$\mu$m may 
also contribute to some extent to this feature. 

\begin{figure}
\epsscale{1.}
\plotone{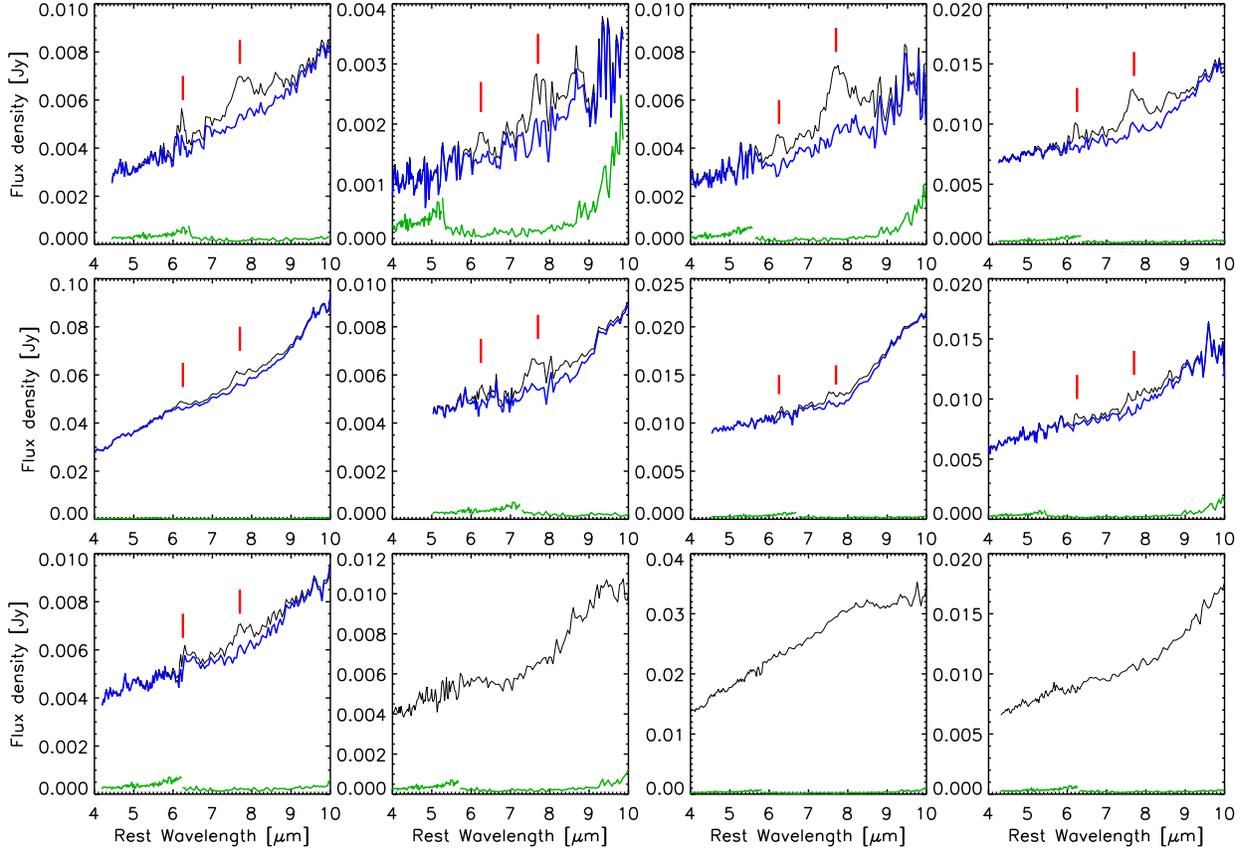}
\caption{Decomposition of IRS spectra. The individual panels are 
shown in the same ordering and scaling as in Fig.~\ref{fig:allhighz}, with the
thin black lines retracing the spectra as observed. Thick blue lines show
the AGN continuum and silicate emission remaining after subtraction of the PAH
component obtained in the decomposition. The green line indicates the 1$\sigma$
noise measured in IRS data obtained with same observing setup and integration 
time but lacking a source (one of the nondetections among the SMGs of 
\citet{valiante07}). While noise from this curve in good agreement with 
expected IRS
sensitivity and measured noise for our continuum-weaker QSOs and gives an 
indication of 
trends with wavelength, it misses possible sources of noise in spectra 
with bright continuum that can arise e.g. due to more 
difficult identification of outlying values. Instead, we conservatively 
present in Table~\ref{tab:results} noise measured in the PAH region of the 
spectra after subtraction of the PAH template. This noise is the RMS after 
subtracting a linear fit to the rest wavelength 6.5--8$\mu$m region (before
the main upturn of silicate emission).}
\label{fig:allhighzdecomp}
\end{figure}

In this paper, we focus on analysis of the PAH emission and its implications 
for star formation in the QSO hosts. Given the S/N of the spectra and the 
variations in slope and shapes of continuum and silicate emission
in AGN, we do not attempt to derive independent fluxes for the various PAH 
features. Informed by the spectra of local QSOs and supported by the average
spectrum of Fig.~\ref{fig:averagehighz}, we quantify the overall 
level of PAH emission by decomposing
the QSO spectra into a scaled and resampled starburst template 
\citep[based on the ISO-SWS spectrum of M82,][]{sturm00} at the
nominal redshift of the QSO, and the underlying AGN emission which 
consists of a dust continuum and silicate emission that is turning up 
beyond $\sim$8$\mu$m. Results of this decomposition are shown in 
Figure~\ref{fig:allhighzdecomp}.
The combination of AGN continuum and silicate emission varies 
smoothly with wavelength over the range of interest. The first signature 
of PAH in such spectra that is already visible at modest S/N then is the 
strongest 7.7$\mu$m feature and its short wavelength steep flank, while its
longer wavelength side is often less pronounced, due to the combined presence
of 8.6$\mu$m PAH and the upturn of silicate emission. The strength of the 
starburst component in the decomposition is adjusted to produce a smooth 
residual AGN emission at the locations of the 6.2 and 7.7$\mu$m features, in 
particular to remove the steep short wavelength flank of the 7.7$\mu$m feature.
Given the S/N of the data we do not attempt to derive PAH feature ratios by 
multi-feature fits, but obtain from the simple template a more robust
characterisation of the overall level of PAH emission which we represent
by the flux of the brightest (7.7$\mu$m) feature. These PAH 7.7$\mu$m fluxes
(Table~\ref{tab:results}) are measured in the scaled starburst template 
resulting from the decomposition, by fitting Lorentzian
features for the PAHs plus an underlying residual continuum.
PAH 7.7$\mu$m fluxes listed in Table~\ref{tab:results} can thus be
directly compared to the results for local QSOs presented by 
\citet{schweitzer06}. 

We list in  Table~\ref{tab:results} the measured noise per resolution 
element over the rest frame 6.5--8$\mu$m region of the actual spectra, 
defined as RMS around a linear fit to the data after subtraction of the 
scaled PAH template and its associated feature structure. By quoting 
the noise measured in this way from the spectra, we implicitly include 
possible noise contributions for stronger continua, that may arise for 
example due to more difficult outlier identification and fringing.
This noise may be conservatively overestimated if the continuum is slightly
curved over this range. Noise $\sim$0.2mJy per spectral sample as obtained
for some of our weaker continuum targets is in good agreement with the
nominal IRS sensitivity for our observing parameters, as implemented
in the Spitzer Science Center SPEC-PET tool.

For the 9 sources for which we report PAH detections in 
Table~\ref{tab:results}, the peak height of PAH emission in the starburst 
component obtained from the decomposition is required to be at least 3 times 
the measured local noise. These significances of 3--13 for the PAH emission 
effectively reflect the significance of the short wavelength 
flank of the 7.7$\mu$m feature. At longer wavelengths, systematic variations 
in shape of underlying continuum and in particular the strength and detailed
shape of the onset of AGN silicate emission at $\geq 8\mu$m start to 
become more and more limiting, preventing reliable use of these wavelengths 
to verify or improve the formal significance. Several 
of the targets (SMMJ04315+102177, KUV0806+4037, HS1002+4400, LBQS1230+1627B) 
show PAH with peak height that is for the individual sources just above
the peak/noise=3 detection limit. In combination with the possibility 
that the mentioned uncertainties on AGN continuum and silicate shape affect 
to a lesser extent also the short flank of the 7.7$\mu$m PAH, we estimate   
that individual PAH fluxes for these weak PAH detections can be uncertain
at the 30--50\%\ level. We caution that they should not be used for 
analyses going significantly beyond the modest accuracy requirements of 
the luminosity analyses presented below.

Upper limits are estimated analogously, for the PAH template scaled to 
a level where the features would remain visibly detectable on top of the 
continuum and the silicate feature. The IRS spectra of the sources with 
PAH nondetections are still consistent 
with the presence of PAH emission at peak feature fluxes similar to the 
sources with detections. 
Again, the main limitation is systematic, due to the strong and structured underlying
continuum and silicate emission in these targets. In the spectrum of 
LBQS 0018-0220, for example, inspection of Fig.~\ref{fig:allhighz} shows that 
the 7-8$\mu$m region of the spectrum, observed with good S/N,  may 
represent either PAH emission partly filling in the region before an 
upturn of silicate emission, or be due to variations in shape/slope of this 
feature, which is strong for LBQS 0018-0220, and the underlying continuum.
Such variations are observed in local AGN \citep[e.g.][]{hao05,schweitzer06,netzer07}.
We list in Table~\ref{tab:results} PAH upper limits considering this possiblity
rather than only reflecting the detector noise.

We will compare below the PAH emission of high-z QSOs with the rest frame 
far-infrared emission. Partly, this is possible by simply using the rest
frame $\sim 250\mu$m continuum that can be obtained from the published
(sub)mm photometry
with minimal extrapolation. For other purposes and for comparison with
the local QSO population we need to extrapolate to the rest frame far-infrared
peak which is constrained by direct observations for very few high redshift 
QSOs only. We do so by assuming that the detected (sub)mm emission reflects
a rest frame T=40K, emissivity index $\beta$=1.5 greybody spectrum. These 
parameters are in reasonable agreement with the dust temperature 
$\sim$36K suggested
for the submillimeter galaxy population by \citet{chapman05} and with
40-60K 
suggested by \cite{beelen06} for high-z QSOs, from the combination of 
350$\mu$m photometry with longer wavelength observations.
For three QSOs overlapping with our sample (KUV 0806+4037, HS 1002+4400, SBS 1408+567), \cite{beelen06} find 32K, 38K,
and 35K respectively, and the FIR-warmer QSOs in their small sample may include
objects with a smaller ratio of host FIR/submm to AGN emission. Similarly, our 
simple extrapolation to rest frame 60$\mu$m for H1413+117 (the Cloverleaf) is 
in good agreement with the observations of \citet{benford99} and 
\citet{roro00}. At redshift z$\sim$2.4, changing the adopted temperature
by $\pm$5K will change the estimated rest frame 60$\mu$m flux by $+$60\% and 
$-$45\%, respectively. This is the dominant source of uncertainty in the estimated
rest frame 60$\mu$m emission, since the original (sub)mm measurements for our IRS sample are
S/N$>$5 by sample definition. Using the T=40K $\beta$=1.5 assumption we extrapolate 
to the rest frame FIR peak of this
component for all our sources, which we express as $\nu L_\nu (60\mu$m) for 
consistency with the local QSO work of \citet{schweitzer06}. These 60$\mu$m 
luminosities are listed in Table~\ref{tab:results}. For sources with 
published independent 850$\mu$m and
1.2mm photometry we have averaged results from the two possible extrapolations.

\section{Discussion}
\subsection{Consistent evidence for luminous starbursts from PAH and rest 
frame far-infrared}

Comparison of the PAH detections and limits to the rest frame far-infrared
emission of the sample QSOs, and comparison to other populations is the key
to testing the hypothesis that both PAH and far-infrared trace intense 
starbursts in the hosts.
We do such comparisons first in directly observed quantities, and then using
PAH luminosities in combination with FIR luminosities derived with the 
greybody extrapolation described above.
 
\begin{figure}
\epsscale{.70}
\plotone{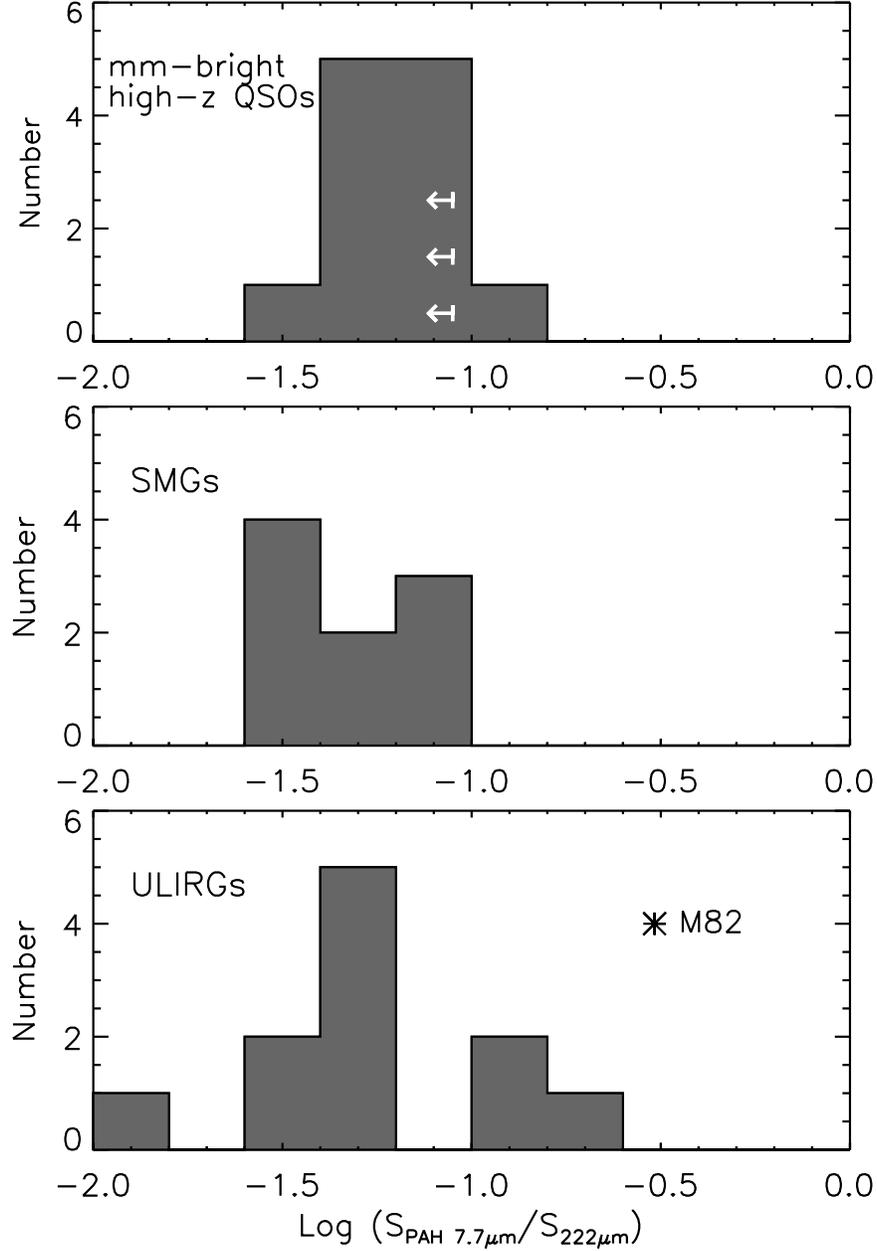}
\caption{Ratio of PAH peak flux density and rest frame 222$\mu$m far-infrared
continuum for the sample QSOs (top panel). z$\sim$2 SMGs and local ULIRGs 
are added for comparison in the lower panels
\citep[SMGs and ULIRGs adopted from][]{valiante07}.}
\label{fig:pah222}
\end{figure}

Fig.~\ref{fig:pah222} shows a comparison of the PAH 7.7$\mu$m peak flux
density with the flux at rest frame wavelength 222$\mu$m. The latter 
corresponds to 
an observed wavelength of 850$\mu$m for z$\sim$2.8, the specific redshift 
chosen for consistency with previous work on SMGs \citep{lutz05,valiante07}. 
This is a simple diagram based on directly observed quantities,
and has minimal sensitivity to SED assumptions for the extrapolation from the 
observed submm/mm point to rest frame 222$\mu$m. 
The distribution of the ratio of 
continuum-subtracted PAH peak flux density and rest frame FIR continuum  
$S_{PAH 7.7\mu m}/S_{222\mu m}$ is 
very similar for our mm-bright QSOs, for the submillimeter galaxy sample
of \citet{valiante07}, and for local ULIRGs. The median 
log($S_{PAH 7.7\mu m}/S_{222\mu m}$) is -1.14, -1.32 and -1.26 for the QSOs, 
SMGs, and ULIRGs, respectively, and the distribution for the QSOs can be 
drawn at $>$25\% probability from the same distribution as either the SMGs 
or the ULIRGs, according to a two-sided Kolomogorov-Smirnov test.

\begin{figure}
\epsscale{.80}
\plotone{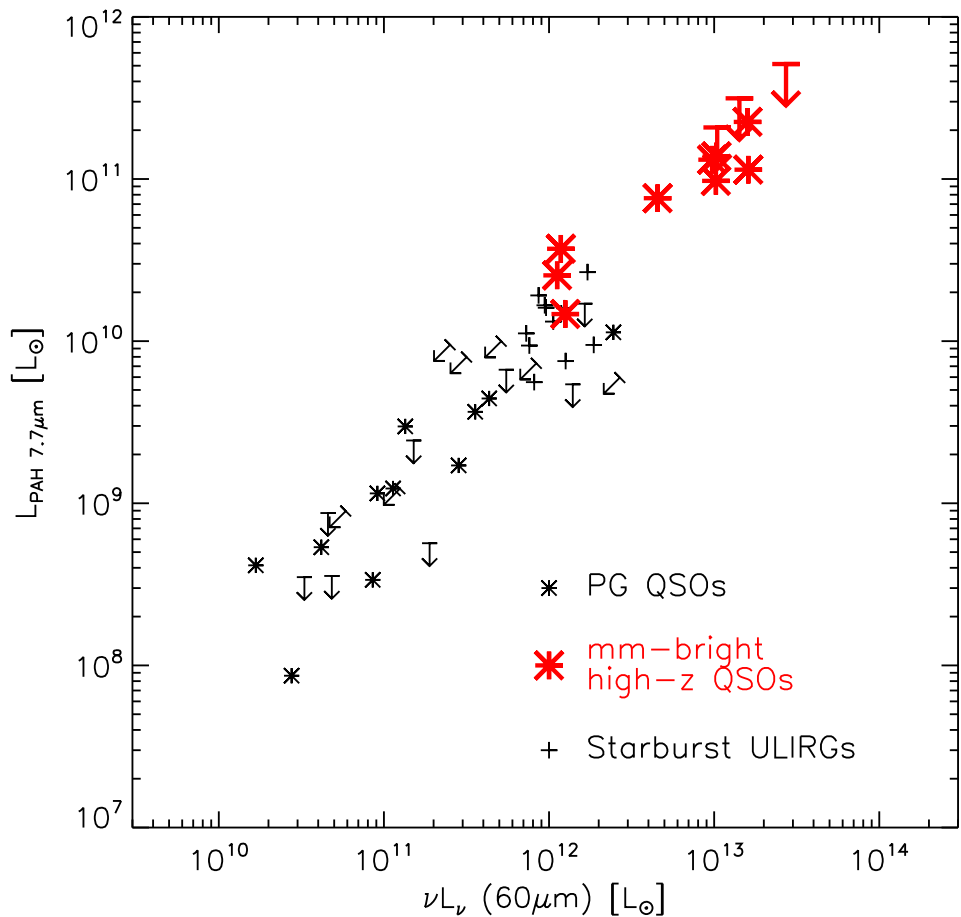}
\caption{Relation of the PAH 7.7$\mu$m luminosity and far-infrared continuum 
luminosities for our sample and for local comparison objects. PG QSOs as 
well as the comparison sample of starbursting ULIRGs are from 
the studies of \citet{schweitzer06} and \citet{netzer07}. For lensed 
high-z objects, 
luminosities have been corrected for the magnifications listed in 
Table~\ref{tab:sample}.}
\label{fig:pahtofir}
\end{figure}

Moving from observed quantities to PAH and far-infrared luminosities,
Fig.~\ref{fig:pahtofir} shows that the mm-bright high-z QSOs extend
the relation between these luminosities found for local PG QSOs by 
\citet{schweitzer06} by an order of
magnitude towards higher luminosities, and that the relation for the QSO 
population crosses the locus for star-forming ULIRGs. The PAH and FIR data
of the mm bright QSOs agree with a scenario where these two tracers are 
jointly originating in 
starbursts in the hosts. These starbursts reach SMG-like luminosities of 
$\sim 10^{13}\Lsun$ in the most luminous (unlensed) targets of our sample. 

Individual PAH-based star formation rates can be derived for our sample and
are listed in Table~\ref{tab:results}. Concerning the calibration
factor applied, the similarity of the PAH/FIR ratio in our sample objects and 
ULIRGs permits 
us to use the ratio L(PAH)/L(FIR)$\sim$ 0.013 derived for the starburst-like 
ULIRGs in 
\citet{schweitzer06}, in combination with the calibration of FIR-based star 
formation rates by \citet{kennicutt98}. Because of the 
changes of PAH/FIR with galaxy properties discussed above, it is worth 
recalling that the resulting 
$SFR\ [\Msun\ year^{-1}]=3.5\times 10^{-42}L_{PAH}\ [erg\ s^{-1}]$ assumes
ULIRG-like physical conditions in the interstellar medium of the galaxy 
observed. PAH based star formation rates for our high-z sample 
range from about 200 to 3000 \Msun\//yr. As was found for
local QSOs and their far-infrared emission \citep{schweitzer06}, use of
the PAH diagnostic provides strong support for massive star forming
events in the mm bright QSOs, at yet an order of magnitude larger 
luminosities. 

Molecular gas CO emission has previously been searched for in six of our 
targets, detecting five (Table~\ref{tab:results}, all gas masses based on 
the conversion factor 0.8 \Msun/(K km s$^{-1}$ pc$^2$) of \citet{downes98} 
and our adopted cosmology). Another target (HE 1104-1805)
has been searched for CO \citep{barvainis02b}, but the nondetection may not be 
meaningful given the narrow bandwidth of this CO search, which may miss the
target CO line for plausible uncertainties of the optical input redshift. 
Consistent with their intense star 
formation, the QSO hosts are gas-rich systems (molecular gas mass of a few
10$^{10}$\Msun). Despite the large gas masses, gas exhaustion times  
M(H$_2$)/SFR are short and range from 10$^7$ to $8\times 10^{7}$ years, where
measured.  

\subsection{From mm-bright QSO1s to the unbiased QSO1 population}

The previous section has discussed properties of the mm-bright
part of the z$\sim$2 luminous QSO1 population only. As an indication of
the significance of this selection, we note that we actually
observed 6 mm-bright QSOs out of the combined samples of \cite{omont03} and
\citet{priddey03}, which in total contain 83 distinct objects. We discuss
the properties of our objects in the context to these two samples, the
1.7$<$z$<$3 objects among the lensed sample of \citet{barvainis02}, 
and the local
PG QSOs, in order to elucidate implications for the complete and unbiased
population of luminous z$\sim$2 QSO 1s.   

\begin{figure}
\epsscale{.80}
\plotone{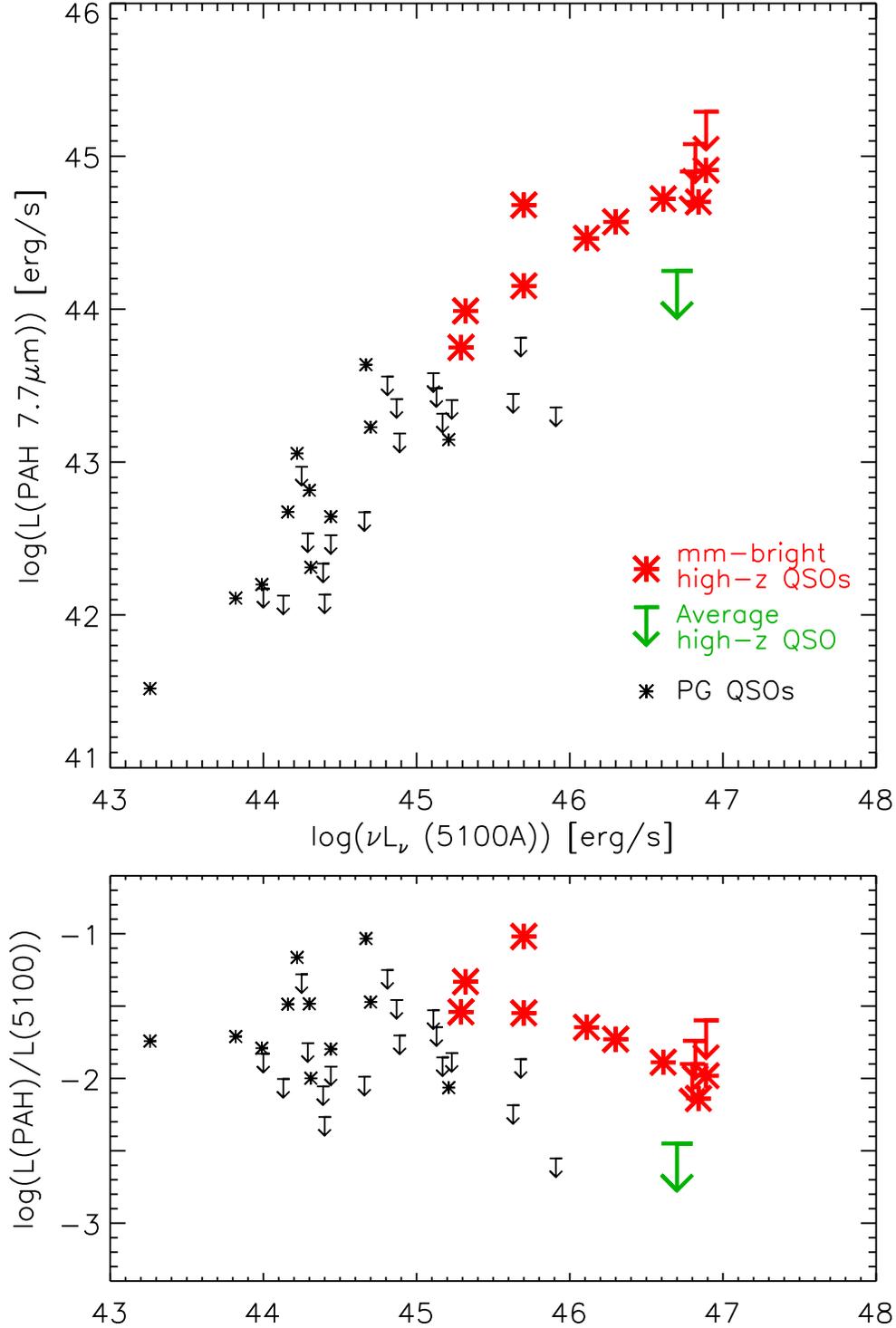}
\caption{Top panel: Relation of PAH 7.7$\mu$m luminosity and AGN 5100\AA\ 
continuum 
luminosity. The large `Average high-z QSO' upper limit symbol refers to 
the limit derived by 
\citet{maiolino07} for the average spectrum of a high redshift QSO sample 
which is not selected for bright (sub)mm emission. Bottom panel: To help 
identifying trends, the same data are shown plotting the ratio of 
PAH 7.7$\mu$m luminosity and AGN 5100\AA\ continuum luminosity as a function
of  AGN 5100\AA\ continuum luminosity.}
\label{fig:pahto5100}
\end{figure}

Figure~\ref{fig:pahto5100} shows the QSO PAH luminosity as a function of
AGN rest frame optical luminosity. The mm bright QSOs clearly extend to
higher luminosity the relation for local PG QSOs already discussed by 
\citet{netzer07}, with a possible saturation (flattening of the relation)
at the largest
AGN luminosities. An extension to individual PAH detections or significant 
limits for the full population including mm-faint QSOs would be extremely 
difficult even with Spitzer capabilities. Substantial insight can already be 
gained, nevertheless,  by adding to Fig.~\ref{fig:pahto5100} the limit on PAH 
emission set by \citet{maiolino07} for the 
{\em average} of a similar redshift QSO1 sample that
is not mm preselected. Unlike their average spectrum, the individual spectra 
from the \citet{maiolino07} sample are of insufficient S/N to put a 
limit on PAH emission 
that is meaningful for our analysis. Compared to mm bright QSOs from our 
sample of similar AGN luminosities, the limit of PAH emission for the 
not mm-biased 
average spectrum is clearly lower, by a factor $\sim$3 compared to our 
PAH detections. The corresponding average SFR limit is 
$\lesssim$700$\Msun\ yr^{-1}$. 

\begin{figure}
\epsscale{.75}
\plotone{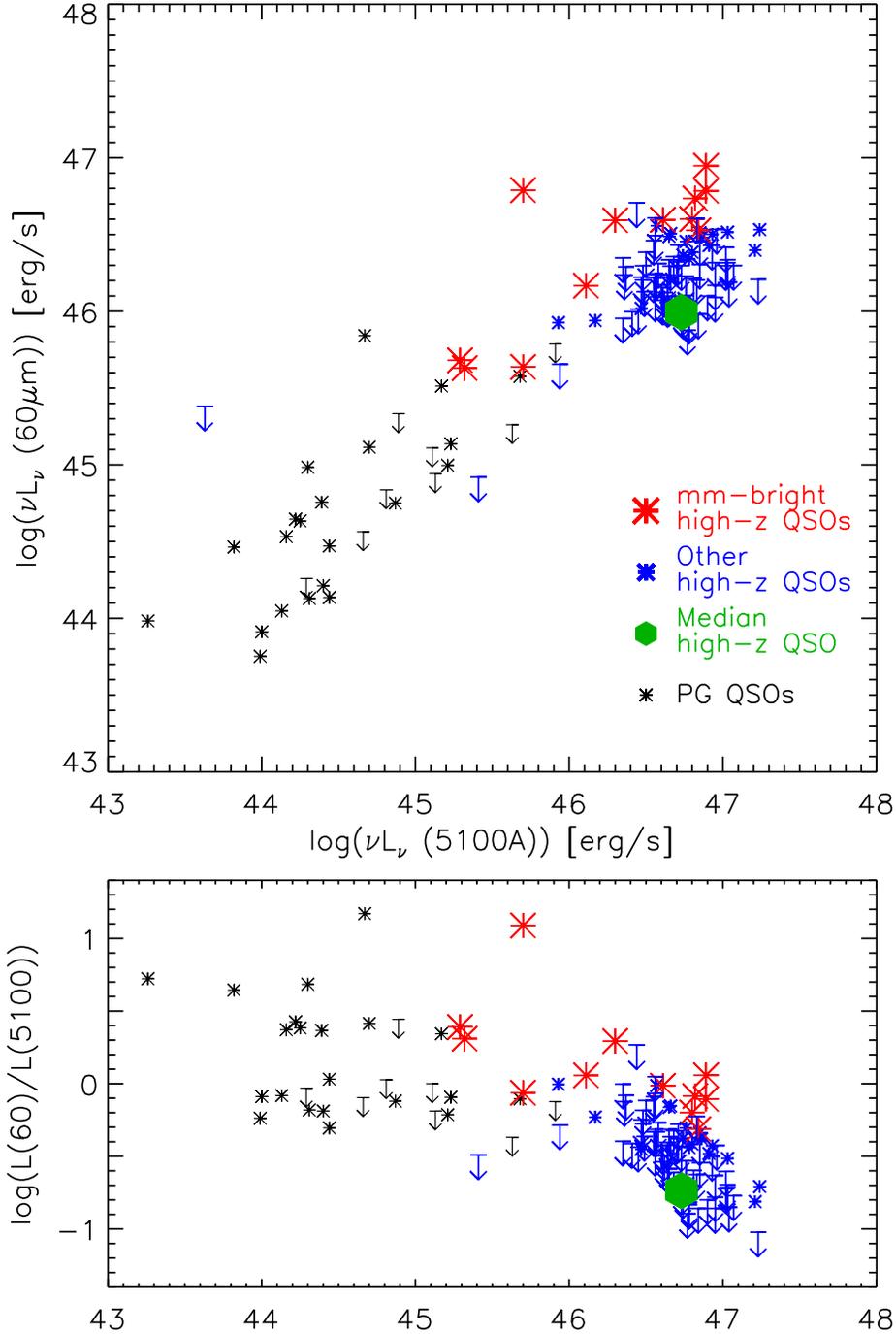}
\caption{Top panel: Relation of 60$\mu$m continuum luminosity and AGN 5100\AA\ 
continuum luminosity. The 60$\mu$m luminosities of high redshift QSOs are 
extrapolated from their (sub)mm photometry using a T=40K $\beta$=1.5 greybody 
approximation. The green hexagon indicates the locus for the median of the 
full z$\sim$2 L$_{5100}>10^{46}$erg\,s$^{-1}$ population. Bottom panel: To 
help identifying trends, the same data are shown plotting the ratio of 
FIR luminosity and AGN 5100\AA\ continuum luminosity as a function
of  AGN 5100\AA\ continuum luminosity. `Other' high-z QSOs refers to objects
in the studies of \citet{omont03}, \citet{priddey03}, and \citet{barvainis02}
that are not part of our mm-bright sample.}
\label{fig:firto5100}
\end{figure}

In an analogous way, Fig.~\ref{fig:firto5100} compares the QSO far-infrared
luminosity, for which a larger number of individual measurements is available,
to the AGN rest frame optical luminosity. This diagram also reminds
that the z$\sim$2  mm-bright QSOs, having L$_{FIR}\sim$L$_{5100}$ i.e.  
L$_{FIR}\sim$0.1 L$_{Bol}$, are far from being extreme infrared excess objects
and resemble in this property the average local QSO. In relation to the
AGN luminosity, the strongest IR emission 
for the high-z QSOs in this diagram is for SMMJ04315+1027, which was 
originally detected at submm wavelengths. In this diagram, the bulk of the 
SMG population would extend from the location of SMMJ04315+1027 to the left,
because of the typically modest luminosity of the AGN they are often hosting
\citep[e.g.][]{alexander05,valiante07}.
As concerns the relation of
mm bright QSOs to the total QSO population, we obtain a similar result as from
the PAH data. Again, the mm-bright QSO studied here extend the relation 
for local QSOs and represent the far-infrared bright
end of a distribution in L$_{FIR}$/L$_{5100}$ among the z$\sim$2 QSOs. 
Considering only the L$_{5100}>10^{46}$erg s$^{-1}$ QSOs, the mm bright
QSOs have median L$_{FIR}$=10$^{46.6}$erg\,s$^{-1}$, a factor of 4 larger 
than the median
for the full population. The specific value of this factor should be 
considered tentative only, since the distribution for the full population 
is based on (sub)mm photometric points of typically low individual 
significance.

The PAH spectroscopy presented for mm bright high-z QSOs here, and for 
local QSOs in \citet{schweitzer06} and \citet{netzer07}, has provided 
strong support for
a star formation origin of the rest frame far-infrared emission in these 
systems. This conclusion has been derived for QSOs with  
L$_{FIR}\sim$L$_{5100}$, however, and needs to be critically reviewed for
the even more AGN dominated objects at the mm-faint end of the
high-z QSO population. In \citet{netzer07}, we have presented a 
host-subtracted intrinsic
AGN SED for local PG QSOs which is very broadly characterised by
$\nu L_\nu$(5100\AA\/)$\sim \nu L_\nu$(6$\mu$m)$\sim 5\times \nu L_\nu$(60$\mu$m), 
with considerable scatter between individual objects. This intrinsic SED drops 
significantly towards the rest frame far-infrared, but in combination with 
the properties of the high-z QSOs (Fig.~\ref{fig:firto5100}) would 
imply that the mm-faint end of the z$\sim$2 QSO population
in fact may contain almost `pure' AGN without strong host star formation. 
The AGN-heated dust could already provide the observed weak rest frame 
far-infrared
emission. A caveat to be noted here originates from recent 
evidence for systematic changes
in the mid-infrared to optical SEDs of luminous high-z QSOs compared
to their local low-luminosity analogs \citep{maiolino07,treister08}, and from 
the possible detection of high-z AGN without hot dust \citep{jiang06}. 
We defer a detailed analysis of the AGN-heated dust in our sample to a future 
paper, but note that the larger PAH equivalent width of our 
average spectrum compared to the local PG average 
(Fig.~\ref{fig:averagehighz}) is in support of these results, 
given that it is combined with similar 
or lower ratio of PAH and 5100\AA\ continuum in our sample compared to locally
(Fig.~\ref{fig:pahto5100}). 
These results
suggest a lower mid-infrared AGN continuum luminosity in relation 
to the optical AGN luminosity, and thus smaller AGN dust 
covering factors. They imply changes of the intrinsic AGN SED for luminous 
high-z QSOs compared to the \citet{netzer07} intrinsic SED for local QSOs,
likely weakening the AGN-related 60$\mu$m emission along with the weakening
of the rest frame mid-infrared AGN dust emission. This decrease of AGN 
rest frame 60$\mu$m emission would reduce again the part of 
the total far-infrared continuum directly ascribed to the AGN and perhaps 
suggest a noticeable host contribution even in mm-faint high-z QSOs. 
Firmly establishing the nature of the weak 
rest frame far-infrared emission at the mm-faint end of the z$\sim$2 QSO 
population will hence require a more complete SED characterisation over
the full rest frame infrared range. 

\subsection{A flattening relation of starburst and AGN luminosity}

For local QSOs with L$_{5100}\sim 10^{44}$ to 10$^{46}$erg\,s$^{-1}$,
we have found clear correlations of the star formation indicators PAH and 
far-infrared continuum with AGN optical luminosity \citep{netzer07}, that
are approximately linear over this luminosity range. The high-z QSOs extend 
the AGN luminosity range by another order of magnitude. The relation of PAH 
(star formation) and AGN luminosity is flattening at  
L$_{5100}> 10^{46}$erg\,s$^{-1}$ (Fig.~\ref{fig:pahto5100}). This is not yet
unambiguous from the comparison of our mm-bright sample alone with the 
local PG QSOs. There may be a slight flattening, and the slope of a 
linear fit in 
Log(L$_{PAH}$) vs. log(L$_{5100}$) is 1.01 for the PAH detections among 
the local PG QSOs, versus 0.92 adding the high-z PAH detections to the
sample, but these changes are not yet significant. The effect becomes clear 
by inclusion of the PAH upper limit which has been derived 
for the sample of \citet{maiolino07} that is not selected to be mm-bright 
(Fig.~\ref{fig:pahto5100}). In agreement with mm continuum 
based studies, our PAH spectroscopy 
does not find evidence for star formation rates above 
3000\Msun yr$^{-1}$ (starburst luminosities much larger than 10$^{13}\Lsun$).
This saturation occurs at luminosities similar to the upper end of the 
luminosity range spanned 
by the submm galaxy population, a limit that is often related to the star 
formation rate in a maximal starburst that is converting much of its gas 
into stars over just a few dynamical times \citep[e.g.][]{tacconi06}. 
In support of this view, a slower than linear increase (or saturation at 
high AGN luminosities) is also inferred for the relation of cold gas 
mass and AGN luminosity, from a synopsis of CO measurements for 
high-z/high-L QSOs \citep{maiolino07b}.

Local PG QSOs are found in a region of the fundamental plane for elliptical
galaxies and spheroids that is hosting 
moderate mass ellipticals with corresponding black hole 
masses $\sim10^8$\Msun\/, and accreting
efficiently at an Eddington ratio $\sim$0.25 \citep{dasyra07}. Relative to the
host star formation rate, their AGN growth rates are more than an order of 
magnitude larger than needed to explain the locally observed BH mass to 
bulge mass relation under the assumption of coeval growth \citep{netzer07}.
The increased L(AGN)/L(SF) for the high-z QSOs and in particular the mm-faint
ones places them in a yet more pronounced phase of AGN growth proceeding much
more rapidly than host growth.
The failure up to now to identify matching extreme $>>10^{14}\Lsun$ star 
forming events also in other galaxies and surveys indicates this is not 
just due to a time delay between AGN and host growth,
but must reflect intrinsically different durations of the AGN growth and host 
growth phases.

\section{Conclusions}

In the context of quantitatively constraining the coexistence of star 
formation and AGN activity in different classes of high redshift galaxies, we 
have used Spitzer-IRS mid-infrared spectroscopy to detect PAH emission
in the spectra of nine out of twelve mm-bright z$\sim$2 radio quiet QSOs,
with optical luminosities L$_{5100}=10^{45}$ to 
10$^{47}$ erg s$^{-1}$.
The detections provide strong support for the presence of intense star 
formation up to $\sim$3000\Msun yr$^{-1}$ in their hosts, that is dominating 
their rest frame far-infrared emission. The typical luminous
z$\sim$2 QSO that is not preselected for mm emission has lower star 
formation rate $<$700\Msun yr$^{-1}$ and the 
mm-faint part of this population may include objects that are AGN dominated
out to rest-frame far-infrared wavelengths, due to luminous AGN but at best
modest level of star formation. Toward the high luminosity and redshift QSOs,
there is a flattening in the relation between starburst and AGN luminosity 
that is observed for local PG QSOs, and no object in our sample has a 
PAH-measured star formation rate above 3000\Msun yr$^{-1}$.

\acknowledgments
This work is based on observations made with the 
\textit{Spitzer Space Telescope},
which is operated by the Jet Propulsion Laboratory, California Institute of
Technology, under a contract with NASA. Support for this work was provided by
NASA under contracts 1287653 and 1287740 (S.V.,O.S.). 

%up to 8 authors, then first et al.

\clearpage

\begin{deluxetable}{llrrrlc}
\tablecolumns{7}
\tablecaption{Sample}
\tablehead{
\colhead{Name}        &
\colhead{z}    &
\colhead{RA}          &
\colhead{DEC}         &
\colhead{$\mu_{L}$}&
\colhead{Ref.}&
\colhead{log($L_{5100}$)}\\
\colhead{}&
\colhead{}&
\colhead{J2000.0}&
\colhead{J2000.0}&
\colhead{}&
\colhead{}&
\colhead{erg/s}
}
\startdata
LBQS 0018-0220  &2.596& 0:21:27.30&$-$2:03:33.0& 1.0&   &46.82\\
HE 0230-2130    &2.162& 2:32:33.10&$-$21:17:26.0&14.5&B02&45.29\\
SMMJ04315+10277 &2.837& 4:13:27.26& +10:27:40.5& 1.3&K03&45.70\\
KUV 0806+4037   &1.795& 8:12:00.54& +40:28:13.0& 1.0&   &46.30\\
RXJ0911.4+0551  &2.793& 9:11:27.61&  +5:50:54.1&20.0&B02&45.32\\
HS 1002+4400    &2.097&10:05:17.43& +43:46:09.3& 1.0&   &46.84\\
HE 1104-1805    &2.305&11:06:33.40&$-$18:21:23.0&10.8&B02&45.70\\
LBQS 1230+1627B &2.735&12:33:10.40& +16:10:52.0& 1.0&   &46.89\\
SBS 1408+567    &2.583&14:09:55.57& +56:28:26.5& 1.0&   &46.89\\
H1413+117       &2.558&14:15:46.27& +11:29:43.4&11.0&V03&46.11\\
HS1611+4719     &2.35 &16:12:39.90& +47:11:57.0& 1.0&   &46.61\\
J164914.9+530316&2.26 &16:49:14.90& +53:03:16.0& 1.0&   &46.89\\
\enddata
\label{tab:sample}
\tablerefs{\ Lensing magnifications $\mu_{L}$ are adopted from: B02: \citet{barvainis02},
 K03: \citet{knudsen03}, V03: \citet{venturini03}. Throughout the paper,
we present fluxes or flux densities as observed, but luminosities, masses, and
star formation rates corrected for these adopted magnifications.}
\end{deluxetable}

\clearpage

\begin{deluxetable}{lrrrrrrr}
\tabletypesize{\scriptsize}
\tablecolumns{7}
\tablecaption{Results and host starburst properties}
\tablehead{
\colhead{Name}        &
\colhead{Peak (PAH7.7)}  &
\colhead{Noise\tablenotemark{a}} & 
\colhead{F (PAH7.7)}     &
\colhead{L (PAH7.7)}    &
\colhead{$\nu L_\nu (60\mu m)$}&
\colhead{SFR (PAH)}&
\colhead{M (H$_2$)}\\
\colhead{}&
\colhead{mJy}&
\colhead{mJy}&
\colhead{$10^{-21}$W\,cm$^{-2}$}&
\colhead{$10^{44}$erg s$^{-1}$}&
\colhead{$10^{44}$erg s$^{-1}$}&
\colhead{\Msun yr$^{-1}$}&
\colhead{10$^{10}$\Msun}
}
\startdata
LBQS 0018-0220  &$<$1.9&0.27&$<$2.3&$<$12.0&543&$<$4200&$<$3.5\tablenotemark{b}\\
HE 0230-2130    &   1.7&0.21&   2.3&    0.6& 48&    197&\\
SMMJ04315+10277 &   0.8&0.20&   0.8&    4.8&615&   1680&13.3\tablenotemark{b}\\
KUV 0806+4037   &   1.1&0.30&   1.7&    3.7&392&   1300&\\
RXJ0911.4+0551  &   2.7&0.21&   3.0&    1.0& 43&    340&0.5\tablenotemark{b}\\
HS 1002+4400    &   1.1&0.35&   1.6&    5.0&337&   1760&\\
HE 1104-1805    &   2.8&0.36&   3.8&    1.4& 43&    496&\\
LBQS 1230+1627B &   1.2&0.28&   1.4&    8.1&609&   2840&2.4\tablenotemark{c}\\
SBS 1408+567    &$<$3.0&0.32&$<$3.6&$<$19.6&888&$<$6840&6.6\tablenotemark{b}\\
H1413+117       &   5.1&0.75&   6.1&    2.9&147&   1020&4.0\tablenotemark{d}\\
HS1611+4719     &   1.0&0.19&   1.2&    5.3&394&   1840&\\
J164914.9+530316&$<$1.5&0.24&$<$2.0&$<$ 7.9&399&$<$2780&\\
\enddata
\tablenotetext{a}{Measured noise per spectral sample in the rest frame 6.5--8$\mu$m range}
\tablenotetext{b}{\citet{hainline04}}
\tablenotetext{c}{\citet{guilloteau99}}
\tablenotetext{d}{\citet{weiss03}}
\label{tab:results}
\end{deluxetable}

\clearpage

\clearpage

\clearpage

\clearpage

\clearpage

\clearpage

\clearpage

\end{document}